\documentclass[12pt]{iopart}
\usepackage{graphicx}
\usepackage{iopams}

\begin{document}

\title[Spin polarization of electron current through a potential barrier]{Spin polarization of electron current through a potential barrier in two-dimensional structures with spin-orbit interaction}

\author{Yurii~Ya.~Tkach, Vladimir~A.~Sablikov and Aleksei~A.~Sukhanov}

\address{Kotel'nikov Institute of Radio Engineering and Electronics,
Russian Academy of Sciences, Fryazino, Moscow District, 141190,
Russia}

\begin{abstract}
We show that an initially unpolarized electron flow acquires spin polarization after passing through a lateral barrier in two-dimensional (2D) system with spin-orbit interaction (SOI) even if the current is directed normally to the barrier. The generated spin current depends on the distance from the barrier. It oscillates with the distance in the vicinity of the barrier and asymptotically reaches a constant value. The most efficient generation of the spin current (with polarization above 50\%) occurs, when the Fermi energy is near the potential barrier maximum. Since the spin current in SOI medium is not unambiguously defined we propose to pass this current from the SOI region into a contacting region without SOI and show, that the spin polarization loss under such transmission can be negligible.
\end{abstract}
\submitto{\JPC}
\maketitle

\section{Introduction}
Generation and manipulation of spin-polarized carriers in semiconductor structures solely by electric methods is a key problem of spintronics ~\cite{Awcshalom,Zutic,Fabian}. One of widely studied approaches to attain this goal is based on using spin-orbit interaction (SOI). The SOI is known to produce the spin polarization of electron current in layered tunnel structures. The effect is caused by the Rashaba SOI at the barrier boundaries of asymmetric structures~\cite{Voskoboynikov} or by the Dresselhause SOI in the barrier bulk~\cite{Perel,Tarasenko,Sandu,Mishra,Fujita}. A general property of such structures is the absence of the spin polarization in the case where the current is directed normally to the barrier. In other words, a current component along a barrier should be created to get a spin current. This limits the capability of these structures to generate spin currents. Recently it has been found that  two-dimensional (2D) structures with a lateral barrier are free of this restriction~\cite{Sablikov}. The electron current passing through the barrier acquires spin polarization, which exceeds 50\% even if the current is directed normally to the barrier. However, in the studied case the SOI exists only inside the barrier. Such structures seem to be hardly realizable, since it is problematic to localize the Rashba SOI within the lateral barrier, especially if the latter is created by gate electrodes.

In the present work the research of Ref.~\cite{Sablikov} is generalized to the case when SOI exists everywhere: in the potential barrier and the surrounding electron gas. We find that the high spin polarization can also be achieved in such structures. However, in this case two important questions arise concerning the definition of the spin current and the existence of equilibrium spin currents in 2D electron gas with SOI~\cite{Rashba}. They provoked recently a wide discussion~\cite{Shi,Son,Sun,Sonin}. As regards the existence of equilibrium spin currents, this question is not essential for the barrier structures considered here for the following reason. The equilibrium spin current is known to be generated only within a narrow energy layer $-E_{so} <E <0$, where $E$ is electron energy, $E_{so}$ is a characteristic energy of SOI~\cite{Rashba}. If the barrier height $U$ considerably exceeds $E_{so}$ the barrier transparency for electrons in this energy layer is negligibly small.

The problem of the spin current definition can be overcome by calculating an observable physical quantity, which is well defined and closely related to the spin flow in SOI medium. This could be a spin current in a normal 2D electron gas (without SOI), which is brought to the contact with the SOI structure under consideration. In other words, it is reasonable to explore a structure in which the spin current generated in the SOI region passes into a normal region where spin current is unambiguously defined. One can say that this region is designed to simulate, at least partially, a measuring process. With this in mind we study the spin current transformation when electrons pass through a contact between SOI and normal 2D regions, and find conditions under which this transformation occurs practically without loss of spin polarization.

Finally we have found, that the barrier in the 2D electron system with the SOI allows one to generate the electron current with spin polarization exceeding 50\% and this spin current can be transferred into a normal 2D electron gas with minimal loss.

\section{Basic wave functions and energy spectrum}
The structure to be studied here is a sheet of 2D electron gas with Rashba SOI separated by a potential barrier into two semiplanes (reservoirs) between which a small voltage $V$ is applied. We are going to find electron and spin currents through the barrier, but begin with a discussion of wave functions for the whole system. The system is described by the Hamiltonian:
\begin{equation}
H=\frac{p_{x}^{2}+p_{y}^{2}}{2m}+\frac{\alpha}{\hbar}(p_{y}\sigma_{x}-p_{x}\sigma_{y})+U(x),
\label{H}
\end{equation}
where $p_{x,y}$ are components of electron momentum, $\alpha$ is the SOI parameter, $\sigma_{x,y}$ the Pauli matrices and $U(x)$ the barrier potential. We consider here a rectangular barrier of high $U$ and width $d$: $U(x)=U$ at $0<x<d$ and $U(x)=0$ at $x<0$ and $d<x$. The effective mass is supposed to be independent of the coordinates.

Wave functions in the barrier and reservoirs are presented in the form of a linear combination of basic eigenfunctions of homogeneous 2D electron gas with SOI
\begin{equation}
\Psi_{{\mathbf k},s}=\sum_{s'}\left[A_{ss',{\mathbf k}}e^{ik_{xs'}x}{\chi_{s'}({\mathbf k})\choose 1}+B_{ss',\bar{\mathbf k}}e^{-ik_{xs'}x}{\chi_{s'}(\bar{\mathbf k})\choose 1}\right]e^{ik_y y},
\label{Psi0}
\end{equation}
where $s$ is the spin index, ${\mathbf k}=(k_{xs},k_y)$ is the wave vector, ${\bar\mathbf k}=(-k_{xs},k_y)$. The wave vector component $k_{xs}$ is different for the barrier and the reservoirs. In addition it depends on the spin. In contrast the component $k_{y}$ is a conserved quantity and hence it is the same for all regions in a given state. The eigenfunctions and energy spectrum of homogeneous electron gas were studied in detail in Ref.~\cite{Sablikov}. The main results which will be used below, are the following. Since the considered system is not translationally invariant in the $x$ direction, the wave vector component $k_{xs}$ can be complex: $k_x=k_x'+ik_x''$. In contrast, $k_y$ is always real. The total spectrum includes three spin-split branches (see Fig.~1).
\begin{figure}
\centerline{\includegraphics[width=8cm]{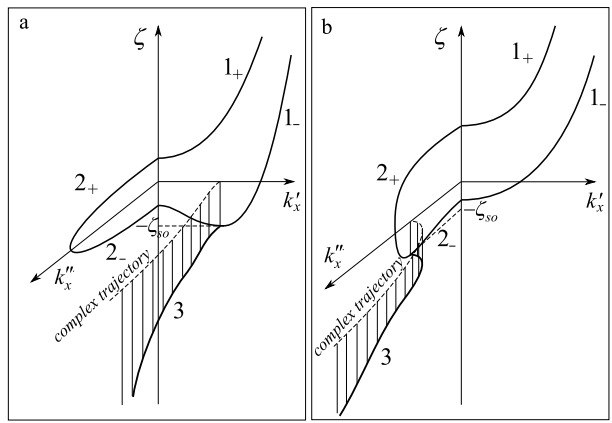}}
\caption{Complex band structure of 2D electron gas with SOI. Energy branches $1_{\pm}, 2_{\pm}$ are shown as functions of $k_x'$ and $k_x''$. Branch 3 is defined along real energy trajectories in a complex plane $(k_x',k_x'')$; only one branch located in the quadrant  $(k_x', k_x''>0)$ is represented. Panels a and b correspond to cases $k_y<a$ and $k_y>a$.}
\end{figure}

1) The first branch $1_{\pm}$ corresponds to propagating states $(k_x''=0)$ with energy:
\begin{equation}
\zeta_{{\mathbf k},s}=-a^{2}+\left(a+s\sqrt{k_y^2+k_x'^2}\right)^2,
\label{zeta1}
\end{equation}
and spin function:
\begin{equation}
\chi_{s}({\mathbf k})=\frac{s(k_y+ik_x')}{\sqrt{k_y^2+k_x'^2}},
\label{chi1}
\end{equation}
where $\zeta_{\mathbf k,s}=2E_{\mathbf k,s}m/\hbar^{2}, E_{\mathbf k,s}$ is the electron energy, $a=m\alpha/\hbar^{2}$ is the characteristic wave vector of the SOI, $\zeta_{so}=a^2$ corresponds to the characteristic energy, $E_{so}=\hbar^2a^2/2m$.

2) The second branch $2_{\pm}$ exists when $k_y\neq0$, in the energy gap between branches $1_+$ and $1_-$. These states decay monotonously with $x$ and hence $k_x'=0$. The energy and spin functions are defined by equations~(\ref{zeta1}) and~(\ref{chi1}), where $k_x'$ must be replaced by $ik_x''$.

3) The third branch lies below two above considered branches, $\zeta_{\mathbf k,s}<-a^{2}$. It is defined for real energy trajectories in the complex plain $(k_x',k_x'')$:
\begin{equation}
k_x'^2k_x''^2+a^2(k_y^2+k_x'^2-k_x''^2)-a^4=0\,.
\label{k}
\end{equation}
The energy and spin functions for this branch are
\begin{equation}
\zeta_{{\mathbf k},s}=-a^2-\frac{k_x'^2k_x''^2}{a^2}\,,
\label{zeta2}
\end{equation}
\begin{equation}
\chi_{s}({\mathbf k})=-a\frac{k_y-k_x''+ik_x'}{a^2+ik_x'k_x''}\,.
\label{chi2}
\end{equation}
Note that at any given energy and $k_{y}$, there are 4 eigenstates. In the case of first and second branches, the different eigenstates correspond to different signs of $s$ and $k_x'$ or $k_x''$. For the third branch the eigenstates differ by signs $(\pm)$ of $k_x'$ and $k_x''$.

\section{Spin polarized current through a barrier}
We now turn to the calculation of electron and spin currents flowing normally to the barrier. For simplicity suppose that the 2D electron reservoirs to the left and right of the barrier are equipotential and the potential difference $V$ is small as compared to all characteristic energies of the system. The electron states contributing to the current are located in the energy interval of $eV$ width near the Fermi energy $E_{F}$. In the $(k_x,k_y)$ space, they occupy two semirings corresponding to electrons with opposite spins (fig.~2).
\begin{figure}
\centerline{\includegraphics[width=8cm]{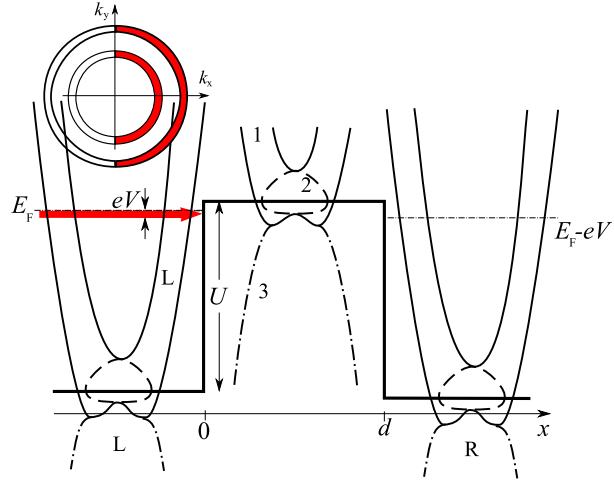}}
\caption{Energy diagram of the barrier structure. Lines $1_{\pm}, 2_{\pm}, 3$ represent the spectrum branches described in the text. In the inset: full semirings in the $(k_x,k_y)$ space which are occupied by electrons contributing to the current.}
\end{figure}
The currents are determined by the summation of partial currents over these states~\cite{Mish,Burkov}. Using variables $\zeta$ and $k_y$ one finds:
\begin{equation}
J(\zeta_F)=\frac{eV}{8\pi^2}\sum_{s}\frac{k_{F,s}}{\sqrt{\zeta_F+a^2}}\int_{-k_{F,s}}^{k_{F,s}}\frac{dk_y}{\sqrt{k_{F,s}^2-k_y^2}} j(\zeta_F,k_y,s),
\label{J}
\end{equation}
where $\zeta_F=2mE_F/\hbar^{2}$, $k_{F,s}$ is defined by the equation:
\begin{equation}
\zeta_F=-a^2+(a+sk_{F,s})^2,
\label{zeta_F}
\end{equation}
and $j(\zeta_F,k_y,s)$ is the partial current in the eigenstate $|\zeta_F,k_y,s\rangle$.

The current $j(\zeta_F,k_y,s)$ is calculated using the wave functions defined in equation~(\ref{Psi0}) as a linear combination of basic eigenfunctions, the set of four eigenfunctions being different for the barrier and reservoirs as well as the spectrum there. The selection of basic eigenfunctions from all three space regions and all spectrum branches to form the total wave function corresponding to a given energy $\zeta$ and transverse momentum $k_y$ is an intricate problem.  Its solution is summarized in the diagram shown in Fig.~3. There are 12 regions on the plane $(\zeta, k_y)$. The regions are bounded by four curves, 1-4, which are determined by the equations:
\begin{equation}
k_y=\sqrt{\zeta+a^2}\mp a\,,
\label{k_y1}
\end{equation}
\begin{equation}
k_y=\sqrt{\zeta-u+a^2}\mp a\,,
\label{k_y2}
\end{equation}
where $u=2mU/\hbar^2$.
\begin{figure}
 \centerline{\includegraphics[width=8cm]{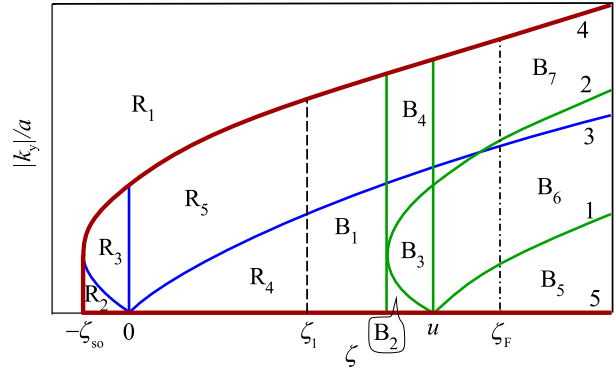}}
\caption{The distribution of the basic eigenfunctions in the plane of parameters ($\zeta, k_y$) for the reservoirs (R) and the barrier (B). The inner borders (lines 1,2,3) determine the regions with different sets of four eigenfunctions. Thick lines 4 and 5 are external borders outside of which no propagating states exist. One more border, closing the region of the states accessible for electrons, is the Fermi energy $\zeta_F$. Dashed line at $\zeta_1$ corresponds to an example considered in the text.}
\end{figure}

In each region a specified set of four eigenfunctions to be used in forming the total wave function is pointed. The list of these regions and corresponding eigenfunction sets for the reservoirs and the barrier are the following:

1) For the reservoirs: $R_1$ is the region without propagating states; $R_2$ contains two waves of spectrum branch $1_-$, which are incident on the barrier, and two waves of branch $1_-$, which are reflected. For brevity we depict this schematically as follows: $R_2$ -- $(\rightrightarrows 1_{-},\leftleftarrows 1_{-}$). The arrows designate the right- and left-moving waves, the number of arrows specifies the number of waves, and the figures behind them indicate the spectrum branches they belong to in accordance with Fig.~1. Using these notation, other regions are imaged as: $R_3$ -- $(\rightarrow1_{-},\leftarrow1_{-},\leftarrow2_{-}$); $R_4$ -- $(\rightarrow1_{-},\rightarrow1_{+}, \leftarrow 1_{-},\leftarrow1_{+}$); $R_5$ -- $(\rightarrow 1_-, \leftarrow 1_-,\leftarrow 2_+$).

2) For the barrier: $B_1$ contains 4 modes of branch 3; $B_2$ -- 4 modes of branch $1_-$; $B_3$ -- 2 modes of $1_-$ and 2 modes of $2_-$; $B_4$ -- 4 modes of $2_-$; $B_5$ -- 4 modes of $1_+$; $B_6$ -- 2 modes of $1_+$ and 2 modes of $2_+$; $B_7$ -- 4 modes of $2_+$.

Fig.~3 helps to find the eigenfunction sets forming the total wave function for a given energy $\zeta_1$. It is needed to draw a vertical line $\zeta=\zeta_1$. The regions, which it crosses, show the eigenfunction sets according to the above list. If this line crosses more than one region, the integration interval in equation (\ref{J}) is to be divided into parts corresponding to its intersection points with internal lines.

As an example, let us describe the tunneling of electrons with energy $E<U-E_{so}$. If an electron falls on the barrier from the left reservoir in the state $|k_{x,s},k_y,s\rangle$, the wave function in this reservoir, $x<0$, is
\begin{equation}
|\psi_{k_{xs},k_y,s}^{(L)}\rangle=|k_{xs},k_y,s\rangle + \sum_{s'}r_{ss'}|-k_{xs'}k_y,s'\rangle\,,
\label{psil}
\end{equation}
where eigenstates $|-k_{xs},k_y,s\rangle$ are those from regions $R_4$ and $R_5$.

The wave function of electrons transmitted to the right reservoir, $x>d$, is
\begin{equation}
|\psi_{k_{xs},k_y,s}^{(R)}\rangle=\sum_{s'}t_{ss'}|k_{xs'},k_y,s'\rangle.
\label{psir}
\end{equation}
Here $r_{ss'}$ and $t_{ss'}$ are the reflection and transmission matrices.

The wave functions in the barrier are formed by the eigenfunctions of ``oscillating'' evanescent states (region $B_1$ in Fig.~3):
\begin{equation}
|\psi_{k_{xs},k_y,s}^{(B)}\rangle=\sum_{\lambda\lambda'}b_{\lambda\lambda'}^s|\lambda K_x',\lambda'K_x'',k_y\rangle,
\label{psib}
\end{equation}
where $K_x'$ and $K_x''$ are the real and imaginary parts of the wave vector $K_x$, $\lambda,\lambda'=\pm 1$.

Matrices $r_{ss'}, t_{ss'}$ and  $b_{\lambda\lambda'}^s$ are defined by an equation set which follows from the boundary conditions~\cite{Sablikov,Govorov,Molen}:
\begin{equation}
\left\{\begin{array}{lll}
\left.\psi \right|_{0_-}^{0_+}=\left.\psi \right|_{d_-}^{d_+}=0\,,\\
\left[ \frac{\partial\psi}{\partial x}+\beta k_y\sigma_z\psi \right]_{0_-}= \left.\frac{\partial\psi}{\partial x}\right|_{0_+},\\
\left.\frac{\partial\psi}{\partial x}\right|_{d_-}=\left[\frac{\partial\psi}{\partial x}-\beta k_y\sigma_z\psi \right]_{d_+}\,.
\end{array} \right.
\label{psi1}
\end{equation}
Here the parameter $\beta=2Ua/eF_{z}$ describes the Rashba SOI caused by a lateral electric field at the edges of the barrier, $F_{z}$ is an electric field normal to the 2D layer. The SOI constant in the boundary condition disappears because the SOI constants are equal all over the sample and the wave functions are continuous at the boundaries. The total equation system for the matrices $t_{ss'}$, $r_{ss'}$ and  $b_{\lambda\lambda'}^s$ is obtained from the boundary conditions for both spin states of incident electrons. Thus, one obtains two systems of 8 equations each. They are to be added by an equation establishing a relation between wave vectors $k_{x,s}$ and $K_x$. This equation follows from the requirement that the energy is the same in the reservoirs and the barrier: $\zeta(k_{xs},k_y,s)=u+\zeta(K_x,k_y)$.

Now we proceed with the calculation of the charge and the spin currents. Using equation~(\ref{psir}), one finds the electron current:
\begin{equation}
j(k_{xs,k_y,}s)=\frac{2\hbar|C|^2}{m}\sum_{s'} [{k_{xs'}}-\frac{ia}{2}(\chi_{s'}-\chi_{s'}^*)]|t_{ss'}|^2\,.
\label{jp}
\end{equation}

The spin current is supposed to be defined by the standard expression~\cite{Mish,Burkov}:
\begin{equation}
j_{s,i}^{j}=\frac{\hbar}{4}\langle\upsilon_{i}\sigma_{j}+\sigma_{j}\upsilon_{i}\rangle,
\label{js}
\end{equation}
where $i=(x,y)$ designates the current components in the plane, $j=(x,y,z)$ designates the spin polarization components, $\upsilon_{i}$ is the electron velocity components.

The calculation of the spin current in the right reservoir for the state $|\psi_{k_{xs},k_y,s}^{(R)}\rangle$ results in the following expressions for the $x$ component:
\begin{equation}
j_{s,x}^{x,y,z}=\frac{\hbar^2|C|^2}{4m} Y_{s,x}^{x,y,z},
\end{equation}
where $Y_{s,x}^{x,y,z}$ has following components:
\begin{eqnarray}
Y_{s,x}^x&=&2\sum_{s'}k_{xs'}|t_{ss'}|^2{\mathrm {Re}}\chi_{s'}+(k_{xs}+k_{x\bar{s}}) \nonumber \\
& &\times[(\chi_s+\chi_{\bar{s}}^*)t_{s,s}t_{s,\bar{s}}^*e^{i(k_{xs}-k_{x\bar{s}})x}+{\mathrm {c.c.}}]/2\,,
\end{eqnarray}
\begin{eqnarray}
Y_{s,x}^y&=&2\sum_{s'}k_{xs'}|t_{ss'}|^2{\mathrm {Im}}\chi_{s'}+i(k_{xs}
+k_{x\bar{s}})[(\chi_{s}- \chi_{\bar{s}}^*)\nonumber \\
& &\times t_{s,s}t_{s,\bar{s}}^*\ e^{i(k_{xs}-k_{x\bar{s}})x}-c.c.]/2- 2a\sum_{s'}|t_{ss'}|^2\nonumber \\
& &-a\left[t_{s,s}t_{s,\bar{s}}^*(\chi_{s}\chi_{\bar{s}}+1)e^{i(k_{xs}-k_{x\bar{s}})x}+{\mathrm {c.c.}}\right],
\end{eqnarray}
\begin{eqnarray}
Y_{s,x}^{z}=(k_{xs}+k_{x\bar{s}})[t_{s}t_{\bar{s}}^*(\chi_{s}+\chi_{\bar{s}}^*)e^{i(k_{xs}-k_{x\bar{s}})x}+{\mathrm {c.c.}}]/2\,.
\end{eqnarray}
Here $\bar{s}$ designates the spin opposite to $s$.
\begin{figure}
\centerline{\includegraphics[width=8cm]{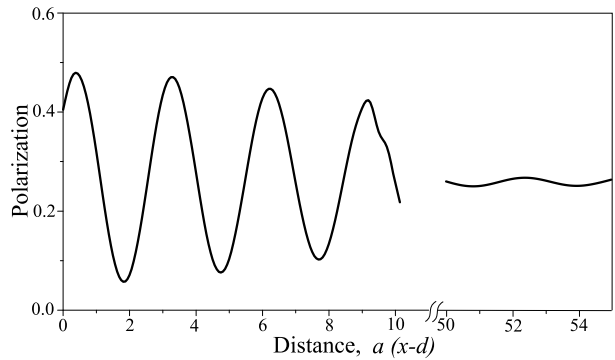}}
\caption{Dependence of the spin polarization of electron current on the distance from barrier. The used parameters are: $E_F/E_{so}=7.99$, $U/E_{so}=9$, $ad=3$, $\beta=0.$}
\end{figure}

The total spin current is:
\begin{eqnarray}
J_{s,x}^{x,y,z}(\zeta_F,x)&=&\frac{eV}{8\pi^2}\sum_s\frac{k_{F,s}}{\sqrt{\zeta_F+a^2}} \nonumber \\
& & \times \int_{-k_{F,s}}^{k_{F,s}}\frac{dk_{y}}{\sqrt{k_{Fs}^2-k_y^2}}j_{s,x}^{x,y,z}(\zeta_F,k_y,s,x)\,.
\label{jsx1}
\end{eqnarray}
Straightforward calculations show that the spin current components with polarization along $x$ and $z$ directions are absent, $J_{s,x}^x=J_{s,x}^z=0$. Only the $y$ component of the spin polarization is present in the spin current $J_{s,x}^y\neq 0$, just as in the case of SOI absence in the reservoirs~\cite{Sablikov}. The spin current depends on the distance from the barrier. Near the barrier, $J_{s,x}^y$ oscillates with a period of about $\pi/a$ around a slowly varying value. The oscillation amplitude decreases with distance and the spin current asymptotically reaches a constant value, as it is shown in Fig.~4. The oscillation is caused by the interference of spin-split propagating states whose wave vectors differ by a value of the order of $a$. An electron incident on the barrier with definite spin appears behind the barrier in a state which is a superposition of wave functions with different chiralities and wave vectors. Their interference results in the spin current oscillations. At large distances from the barrier the interference pattern is smeared because the partial spin current oscillations lose their coherence due to the dispersion of longitudinal wave vectors $k_x$ of incident electrons. The asymptotic behavior of the spin current can be presented as
\begin{equation*}
J_{s,x}^y(\zeta_F,x)\simeq J_{s,x}^y(\zeta_F,\infty)+A(\zeta_F)\frac{\cos[2ax\!+\!\varphi(\zeta_F)]}{\sqrt{x}}\,.
\end{equation*}

The degree of current spin polarization is quantitatively described by the spin-to-charge current ratio:
\begin{equation}
P(\zeta_F)=\frac{2}{\hbar}\frac{J_{s,x}^y(\zeta_F,\infty)}{J(\zeta_F)}\,.
\label{P_s}
\end{equation}
The polarization $P(\zeta_F)$ calculated as a function of the Fermi energy for two thicknesses of the barrier $d$ is presented in Fig.~5. The largest spin polarization is seen to arise when the Fermi level lies close to the barrier maximum in an energy interval of the order of several $E_{so}$. This dependence is similar to that of the case where the SOI is absent in reservoirs~\cite{Sablikov}. This polarization exists at large distances from the barrier. An essential point is that due to the oscillation in the vicinity of the barrier, $P$ can by higher or lower that the asymptotic value shown in Fig.~5.

The obtained results depend only weakly on the parameter $\beta$, describing the interface SOI. With increasing $\beta$ in the range 0-0.1, the general view of the $P(\zeta_F)$ dependence remains unchanged, but the degree of spin current polarization insignificantly increases. So, $P(\zeta_F)$ increases by 7\% as $\beta$ changes from 0 to 0.1.

\begin{figure}
\centerline{\includegraphics[width=8cm]{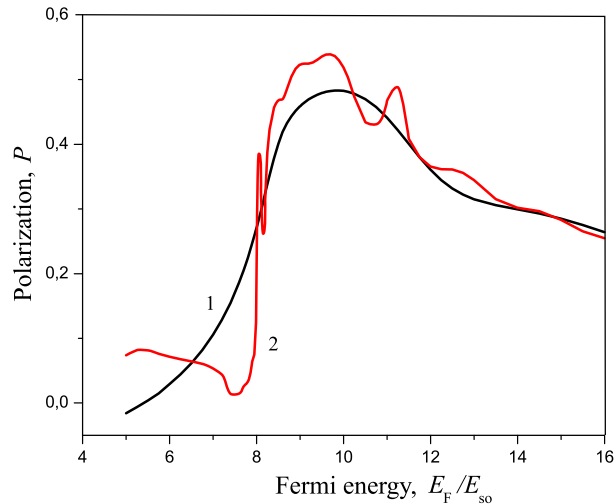}}
\caption{Spin polarization of the current as a function of Fermi energy for two barrier thicknesses: $ad=3$ (line 1), $ad=8$  (line 2). The parameters used are $U=9E_{so}$, $\beta=0$.}
\end{figure}

\section{Spin current transformation in the contact of SOI and normal regions}
In 2D electron gas with SOI, the spin current is known to be a nonconserved quantity and therefore its definition is somewhat arbitrary. For this reason an important question arises of what quantity is really measurable. In this paper we propose to transfer the spin current from the SOI system into a normal 2D electron gas without SOI where the spin current is well defined and measurable~\cite{Hirch,Kato,Valenzuela,Liu}. To carry out this transformation a normal region should be brought into a lateral contact with the SOI system considered before. Thus, it is reasonable to extend the discussed system by adding a contact with a normal region, which simulates (at least partially) a measuring device. The key problem to be solved is to find out how the spin current is transformed while passing through this contact.

The problem is stated as follows. Let a monoenergetic electron flow is incident from the SOI region upon the sharp boundary with a normal region. The spin polarization of incident electrons is determined by a non-equilibrium occupancy of spin states at the Fermi level, which is characterized by distribution functions of the states with positive and negative chiralities, $f_+({\mathbf k}_+)$ and $f_-({\mathbf k}_-)$, with ${\mathbf k}_{\pm}$ being the Fermi wave vectors for the spin-split subbanbs. One needs to calculate the output spin current in the normal region as a function of the spin polarization of the incident current. This problem is solved by the same way as in the previous section. Therefore, we describe below the key results without going into details.

Let us consider a simplified case where the distribution functions $f_+({\mathbf k}_+)$, $f_-({\mathbf k}_-)$ are nonzero only for the states with positive velocity and do not depend on the momentum in this sector of the Fermi surface, i.e. $f_{\pm}({\mathbf k})=f_{\pm}\theta(k_x)$. The ratio of spin-subband populations determines the degree of the spin polarization of the incident electron flow. It is easy to show that
\begin{equation}
\label{P_in}
 P^{(in)}=\frac{\pi}{4}\frac{\zeta_F}{\sqrt{\zeta_F+a^2}}\frac{f_+-f_-}{k_+f_++k_-f_-}\,,
\end{equation}
where
\begin{equation*}
 k_{\pm}=\mp a+\sqrt{\zeta_F+a^2}\,,
\end{equation*}
the spin polarization being directed along $y$ axis.

Note, that far from the contact in the SOI region the spin current does not depend on the coordinate $x$, since electrons occupy the states with well defined spin. Near to the contact, but before it, the situation changes essentially because the electrons having been reflected from the contact find themselves in a superposition of states with different spin. This results in the interference pattern in the spatial distribution of the spin current density similar to that shown in Fig.~4. Behind the contact, in the normal region, the spin current does not depend on the coordinate since it is a conserved quantity.

\begin{figure}
\centerline{\includegraphics[width=8cm]{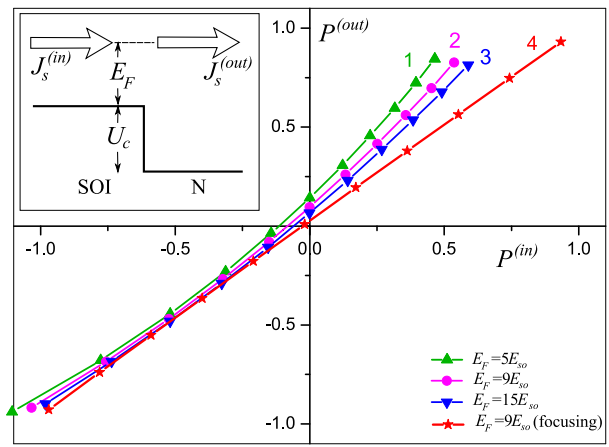}}
\caption{Dependence of the output spin polarization $P^{(out)}$ in the normal region on the input polarization $P^{(in)}$ in the SOI region in the case of $U_c=0$ for different Fermi energies (lines 1,2,3). Line 4 shows the result in the case case, when the distribution function of incident electrons fills the sector ($k_x>0, |k_y|\le 2a$) on Fermi surface. The symbols mark the points, in which $(f_+-f_-)/(f_++f_-)$ increases from -1.0 to 1.0 by steps 0.2, moving from left to right. Inset: the energy diagram of the structure.}
\end{figure}

The spin polarization of the output current $P^{(out)}$ is defined similarly to equation (\ref{P_s}) as the ratio of the transmitted spin current to the particle current. Of interest is the relation between the output polarization, $P^{(out)}$, and the input one, $P^{(in)}$. The input polarization is changed by varying the spin-subband population according to equation (\ref{P_in}). We calculate the output and input polarizations while varying $(f_+-f_-)/(f_++f_-)$ to find a dependence of $P^{out}$ on $P^{in}$. This dependence is determined by the Fermi energy $E_F$ and the potential step height $U_c$ at the contact between the SOI and normal regions. We find that the efficiency of the spin current transformation when transferring through the contact increases with the decrease in $U_c$. This means that the scattering on the contact conributes to the output polarization. The dependencies of $P^{(out)}$ on $ P^{(in)}$ are shown in Fig.~6 for the most favorable case when $U_c=0$. They are nearly linear. Thus, if $E_{F}\gg E_{so},U_c$, the spin current passes from SOI region to normal electron gas practically without polarization loss.

Let us address to the problem of the spin polarization of the electron current through a barrier studied in the previous section.  The largest polarization is reached at $E_F\sim U$. Therefore, if $U\gg E_{so},U_c$, the spin is transferred into the normal 2D gas almost completely even if the distribution functions are uniformly smeared over the semi-circle as in the calculation of this section (see lines 1-3 in Fig.6). In reality, the spin transfer efficiency is higher since the distribution function of transmitted electrons $f_{\pm}({\mathbf k})$ is strongly non-uniform over the azimuthal angle. This occurs because the probability for an electron to pass through the potential barrier decreases with increasing $|k_y|$. One can say, that electrons are focused by the barrier near to $x$ axis. If the energy is close to the barrier top, the characteristic scale of the decrease of $f_{\pm}({\mathbf k})$ with $k_y$ is of the order of $2a$. We model this situation by calculating the polarization of the output current in the case where the states in sector $k_x>0$, $|k_y|\le 2a$ are only filled. The result is presented by line 4 in Fig.~6. Of course, the scattering processes in the bulk reduce the spin polarization because they cause the distribution function to be more isotropic.

\section{Conclusions}
Electron transport through a lateral potential barrier in 2D system with SOI produces the considerable spin polarization of the current, with the spin being directed perpendicularly to the current. Behind the barrier the outgoing spin current depends on the distance in an oscillatory manner, but at sufficiently large distance from it the oscillations decay and the spin current reaches a constant value. The most effective generation of the spin current occurs when the Fermi energy is close to the top of the potential barrier. The maximum degree of polarization at the distance far from the barrier exceeds 50\%. The spin current generated in the 2D electron gas with SOI can be successfully transmitted to a contacting normal 2D electron gas where the spin current is unambiguously defined. The spin polarization loss occurring while electrons pass from the SOI region to the normal electron gas is negligible if the contact potential step and the characteristic SOI energy are small compared to the barrier height and the Fermi energy.

\section*{Acknowledgments}
This work was supported by Russian Foundation for Basic Research (project No. 08-02-00777) and Russian Academy of Sciences (programs ``Basic foundations of nanotechnologies and nanomaterials'' and ``Strongly correlated electrons in solids and structures'').


\section*{References}
\begin{thebibliography}{20}

\bibitem{Awcshalom}
Awcshalom D D, Loss D and Samarth N (ed) 2002 Semiconductor Spintronics and Quantum Computation in the series Nanoscience and Technology (Berlin:Springer)

\bibitem{Zutic}
Zutic J, Fabian J and Sarma S D 2004 Rev. Mod. Phys. \textbf{76}  323

\bibitem{Fabian}
Fabian J, Matos-Abiague A, Ertler C, Stano P and Zutic I 2007 Acta Physica Slovaca \textbf{57}  565

\bibitem{Voskoboynikov}
Voskoboynikov A, Liu S S and  Lee C P 1998 Phys. Rev. B \textbf{58} 15397 ; 1999 Phys. Rev. B \textbf{59} 12514

\bibitem{Perel}
Perel' V I,  Tarasenko S A,  Yassievich I N,  Ganichev S D,  Bel'kov V V and  Prettl W 2003 Phys. Rev. B \textbf{67}, 201304

\bibitem{Tarasenko}
 Tarasenko S A,  Perel' V I, and  Yassievich I N 2004 Phys. Rev.Lett. \textbf{93} 056601

\bibitem{Sandu}
Sandu T, Chantis A, and Iftimie R 2006 Phys. Rev. B \textbf{73} 075313

\bibitem{Mishra}
Mishra S, Thulasi S, and Satpathy S 2005 Phys. Rev. B \textbf{72} 195347

\bibitem{Fujita}
Fujita T, Jalil M B A and Tan S G 2008 J.Phys.: Condens. Matter \textbf{20} 115206

\bibitem{Sablikov}
Sablikov V A and Tkach Yu Ya 2007 Phys. Rev. B \textbf{76} 245321

\bibitem{Rashba}
Rashba E I 2003 Phys. Rev. B \textbf{68} 241315(R)

\bibitem{Shi}
Shi J, Zhang P, Xiao D and  Niu Q 2006 Phys. Rev. Lett. \textbf{96} 076604

\bibitem{Son}
Sonin E B 2007 Phys. Rev. B \textbf{76} 033306

\bibitem{Sun}
Sun Q -F, Xie X C and Wang J 2007 Phys. Rev. Lett. \textbf{98} 196801; 2008 Phys.Rev. B \textbf{77} 035327

\bibitem{Sonin}
Sonin E B 2007 Phys. Rev. Lett. \textbf{99} 266602

\bibitem{Mish}
Mishchenko E G and Halperin B I 2003 Phys. Rev. B \textbf{68} 045317

\bibitem{Burkov}
Burkov A A, N\'u\~nez A S and MacDonald A H 2004 Phys. Rev. B \textbf{70} 155308

\bibitem{Govorov}
Govorov A O, Kalameitsev A V and  Dulka J P 2004 Phys. Rev.B \textbf{70}  245310

\bibitem{Molen}
Molenkamp L W, Schmidt G and Bauer G E W 2001 Phys. Rev. B \textbf{64} 121202(R)

\bibitem{Hirch}
Hirsch J E 1999  Phys. Rev. Lett. \textbf{83} 1834

\bibitem{Kato}
Kato Y K, Myers R S, Gossard A C and Awschalom D D 2004 Science  \textbf{306} 1910

\bibitem{Valenzuela}
 Valenzuela S O and Tinkham M 2006 Nature \textbf{442} 176

\bibitem{Liu}
Liu J-T and Chang K 2008 Phys. Rev. B \textbf{78} 113304

\end{thebibliography}
\end{document}